\pdfoutput=1
\documentclass[11pt]{article}
\usepackage{graphicx}
\usepackage{amssymb}

\newtheorem{theorem}{Theorem}
\newtheorem{corollary}[theorem]{Corollary}

\title{A Linear-Time Approximation Algorithm for Rotation Distance}
\author{Sean Cleary\thanks{
	Department of Mathematics,
	City College of New York,
	City University of New York,  New York, NY 10031,
	{\tt cleary@sci.ccny.cuny.edu}.  
	Partial funding provided by NSF \#0811002.
}
\and 
Katherine St.~John\thanks{
	Department of Mathematics \& Computer Science,
	Lehman College \& the Graduate Center,
        	City University of New York, Bronx, NY 10468,
         {\tt stjohn@lehman.cuny.edu}.
         Partial funding provided by NSF  \#0513660.
}
}

\begin{document}
\maketitle

\begin{abstract}
Rotation distance between rooted binary trees measures the 
number of simple operations it takes to transform one
tree into another.  There are no known polynomial-time
algorithms for computing rotation distance.
We give an efficient, linear-time approximation algorithm, which estimates the rotation distance, 
within
a provable factor of 2, between ordered rooted binary trees. 
\end{abstract}

\section{Introduction}

\begin{figure}
\begin{center}
\includegraphics[height=2.0in]{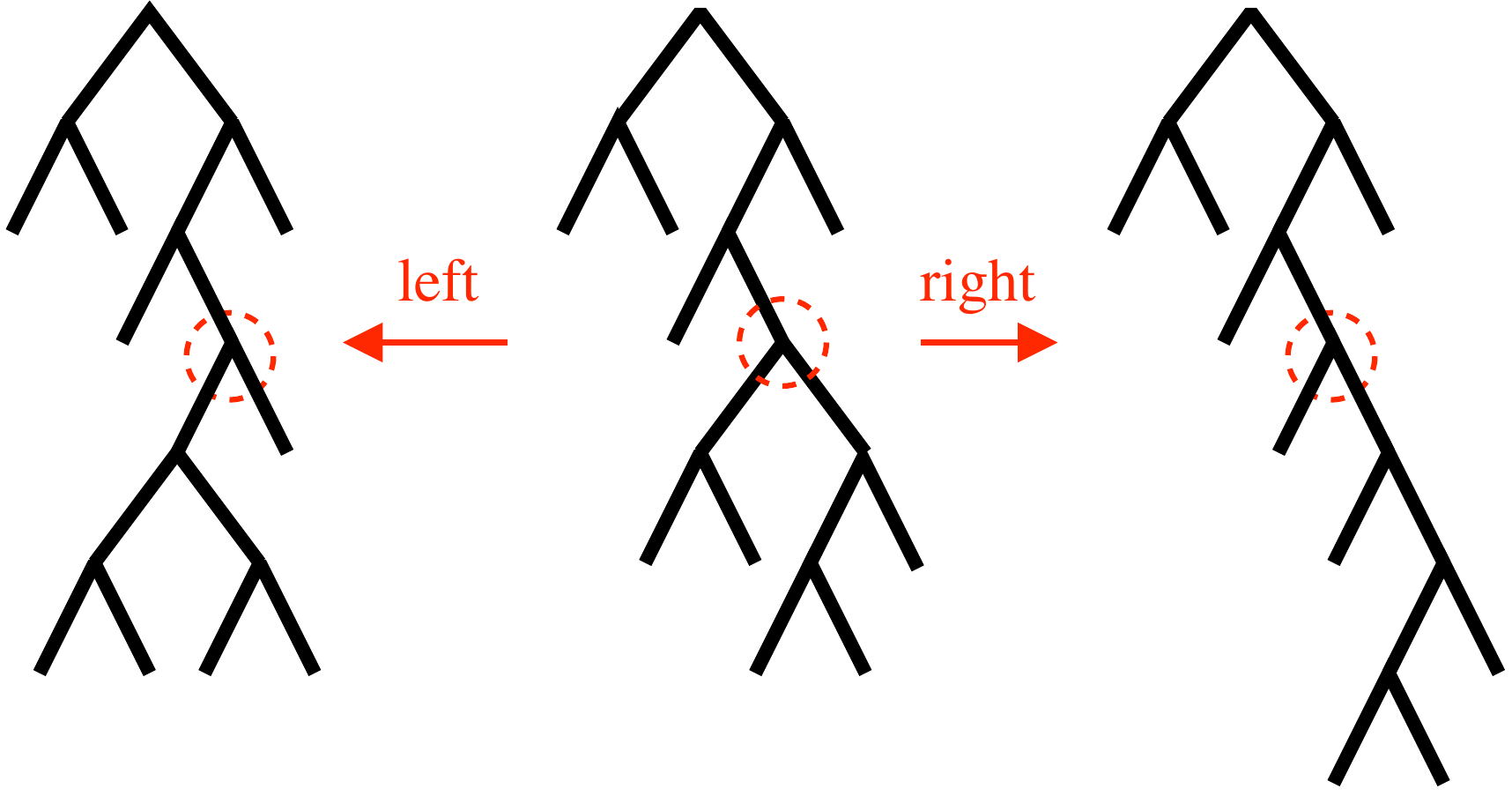}
\end{center}
\caption{\small A (right)
rotation at a node consists of rotating the right child of the left child of the 
node to the right child of the node.  A left rotation is defined similarly by
moving the left child of the right child of the node to the left child of the node.
The circled node in the middle tree has been rotated right to yield the
tree on the right, and similarly rotated left to yield the tree on the left.  }
\label{rot}
\end{figure}

Binary search trees are a fundamental data structure 
for storing and retrieving information \cite{clr}. Roughly,
a binary search tree is a rooted binary tree where the 
nodes are ordered ``left to right.''
The potential efficiency of storing and retrieving information in
binary search trees depends on their height and balance.
Rotations provide a simple mechanism for ``balancing''
binary search trees while preserving their underlying
order (see Figure~\ref{rot}).
There has been a great deal of work on estimating, 
bounding and computing rotation distances. By rotating to right caterpillar trees,
Culik and Wood \cite{cw} gave an immediate upper bound of $2n-2$ for the distance between two trees with $n$ interior nodes. In elegant work using methods of hyperbolic volume, Sleator, Tarjan, and Thurston \cite{stt} showed not only that $2n-6$ is an upper bound for $n \geq 11$, but furthermore that for all very large $n$, that bound is realized.  In remarkable recent work, Dehornoy \cite{dehornoy} gave concrete examples illustrating that the lower bound is at least $2n-O(\sqrt{n})$ for all $n$. There are no known polynomial-time algorithms for computing
rotation distance, though there are polynomial-time estimation algorithms of Pallo \cite{pallo}, Pallo and Baril \cite{barilpallo},
and Rogers \cite{rogers}.  
 Baril and Pallo \cite{barilpallo} use computational
experimental evidence to show that a large fraction of their estimates are within a factor of 2 of the rotation distance.
The problem has been recently shown
to be fixed-parameter tractable in the parameter, $k$, the
distance \cite{cs08}.
Li and Zhang \cite{li1998} give a polynomial time 
approximation algorithm for the equivalent diagonal
flip distance with approximation ratio of  
almost 1.97.\footnote{The exact ratio is bounded by the
maximum number of diagonals, $d$, allowed at any
vertex, and is $2- \frac{2}{4(d-1)(d+6)+1}$.}

In this short note, we give a linear time approximation algorithm with 
an approximation ratio of 2,  improving the running time at the very modest expense of approximation ratio.
This is accomplished by showing the distance between 
the trees is bounded below by $n-e-1$ and above 
by $2(n-e-1)$ where 
$n$ is the number of internal nodes and e is the number
of edges in common in the reduced trees.
The number of common edges is equivalent to
Robinson-Foulds distance, widely used in phylogenetic settings, which Day \cite{day} calculates in 
linear time.

\section{Background}
\label{background-section}

We consider ordered, rooted binary trees with $n$ interior nodes and
where each interior node has two children.  Such trees are
commonly called {\em extended binary trees} \cite{knuth3}.
In the following, {\em tree} refers to such a tree with an ordering on the leaves, {\em node} refers
to an interior node, and {\em leaf} refers to a non-interior node.
Our trees will have $n+1$ leaves numbered in left-to-right order
from 1 to $n+1$.  The size of a tree will be the number of internal nodes it
contains.   Each internal edge in a tree separates the leaves into two connected sets upon removal,
and a pair of edges $e_1$ in $S$ and $e_2$ in $T$ form a {\em common edge pair} if their removal
in their respective trees gives the same partitions on the leaves.  In that case, we say that $S$ and $T$
have a {\em common} edge.

Right rotation at a node of a rooted binary tree is defined as a simple change to
$T$ as in
Figure~\ref{rot}, taking the middle tree to the right-hand one. Left rotation at a node is
the natural inverse operation.
The
{\em rotation distance}
$d_R(S,T)$ between two rooted binary trees
$S$ and
$T$ with the same number of leaves is the minimum number of rotations
needed to transform $S$ to $T$.

The specific instance of the rotation distance problem we address is:
\begin{quote}
{\sc Rotation Distance:}\\
{\sc Input:} Two rooted ordered trees, $S$ and $T$ on 
$n$ internal nodes,\\
{\sc Question:} Calculate the rotation distance between them, $d_R(S,T)$.
\end{quote}

Finding a sequence of rotations which accomplish the transformation gives only an upper bound.  
The general difficulty of computing rotation distance comes from the lower bound. 

\section{Approximation Algorithm}
\label{resultsSection}

We first show that the rotation distance is bounded by the
number of edges that differ between the trees.  From this,
the approximation result follows easily.

\begin{theorem}  Let $S$ and $T$ be two distinct ordered rooted trees with the same number of leaves.  Let $n$ be the
number of internal nodes and $e$ the number of common edges
for $S$ and $T$.  Then, 
$$n-e-1 \leq d_R(S,T) \leq 2(n-e-1)$$    
\label{boundThm}
\end{theorem}

{\em Proof:}
The lower bound follows from two simple observations. First, if we use a single
rotation to transform $T_1$ to $T_2$, all but one of the internal edges in each tree
is common with the other tree.  Second,  every
internal
edge of $S$ that is not common with an internal edge of $T$ needs a rotation
(possibly more than one) to transform it to an edge in common in $T$.
The number of internal edges  occurring only in $S$ is 
$n-e-1$ and thus, is also a simple lower bound.

For the upper bound, we use two facts from past work on 
rotation distance.  We first let $(S_{1},T_{1})$, $(S_{2},T_{2})$,
$\ldots$, $(S_{e+1},T_{e+1})$ be the resulting tree pairs from 
removing the $e$ edges $S$ and $T$ have in common, where we insert placeholder leaves to
preserve the extended binary tree property.  Let
$n_i$ be the size of tree $S_i$ for $i=1,2,\ldots,e+1$. 
The first is the observation
of Sleator {\em et al.} \cite{stt} used before:  the rotation distance of the original tree pair $(S,T)$ with a common 
edge is the sum of the rotation distances of the two tree pairs ``above'' and
``below'' the common edge.  Extending this to $e$ edges in 
common between $S$ and $T$, we have 
$$
d_R(S,T) = \sum_{i=1}^{e+1} d(S_{i},T_{i}) \leq \sum_{i=1}^{e+1} 2n_i -2 = 2n - 2 (e+1) = 2(n-e-1) 
$$
The inequality follows by the initial bound of $2n-2$ on rotation distance between trees with $n$ internal nodes of Culik and Wood  \cite{cw}.

Thus, $n-e-1 \leq d_R(S,T) \leq 2(n-e-1)$.
\hfill$\Box$

We note that using the sharper bound of $2n-6$ for $n>12$ from Sleator, Tarjan and Thurston \cite{stt} together with the table of
distances for $n \leq 12$ can improve this slightly still further.

These reduction rules and counting the number
of common edges can be carried out in
linear-time \cite{approx,day},  yielding the corollary:

\begin{corollary}  Let $S$ and $T$ be ordered rooted
trees with $n$ internal nodes.  A $2$-approximation of their rotation distance
can be calculated in linear time.
\end{corollary}

{\em Proof:}
 Let $S$ and $T$ be two distinct ordered rooted
$n$-leaf trees.  Let $n$ be the
number of internal nodes and $e$ the number of edges in common
for $S$ and $T$.  Then, by Theorem~\ref{boundThm},
$n-e-1 \leq d_R(S,T) \leq 2(n-e-1)$.   Since this is within a linear factor 2 from  both bounds, 
we have the desired approximation.
\hfill$\Box$

We note that this algorithm not only approximates rotation distance, it gives a sequence of rotations which realize the upper bound of the approximation, again in linear time.  The approximation algorithm uses the Culik-Wood bound on potentially several pieces.  On each piece, the
$2n-2$ bound comes from rotating each internal node which is not on the right side of the tree to obtain a right caterpillar, and then rotating the caterpillar to obtain the desired tree.  This can be
accomplish simply in linear time.
\small
\bibliographystyle{plain}

\begin{thebibliography}{10}

\bibitem{barilpallo}
Jean-Luc Baril and Jean-Marcel Pallo.
\newblock Efficient lower and upper bounds of the diagonal-flip distance
  between triangulations.
\newblock {\em Information Processing Letters}, 100(4):131--136, 2006.

\bibitem{approx}
Maria~Luisa Bonet, Katherine St.~John, Ruchi Mahindru, and Nina Amenta.
\newblock Approximating subtree distances between phylogenies.
\newblock {\em Journal of Computational Biology}, 13(8):1419--1434
  (electronic), 2006.

\bibitem{cs08}
Sean Cleary and Katherine~St. John.
\newblock Rotation distance is fixed parameter tractable.
\newblock 109:918--922, 2009.

\bibitem{clr}
T.H. Corman, C.E. Leiserson, and R.L. Rivest.
\newblock {\em Introduction to Algorithms}.
\newblock McGraw-Hill, 1990.

\bibitem{cw}
Karel Culik~II and Derick Wood.
\newblock A note on some tree similarity measures.
\newblock {\em Information Processing Letters}, 15(1):39--42, 1982.

\bibitem{day}
W.~H.~E. Day.
\newblock Optimal algorithms for comparing trees with labeled leaves.
\newblock {\em Journal of Classification}, 2:7--28, 1985.

\bibitem{dehornoy}
Patrick Dehornoy.
\newblock On the rotation distance between binary trees.
\newblock Preprint, arXiv:math.CO/0901.2557.

\bibitem{knuth3}
Donald~E. Knuth.
\newblock {\em The {A}rt of {C}omputer {P}rogramming. {V}olume 3}.
\newblock Addison-Wesley, Reading, Mass, 1973.
\newblock Sorting and searching.

\bibitem{li1998}
Ming Li and Louxin Zhang.
\newblock Better approximation of diagonal-flip transformation and rotation
  transformation.
\newblock In {\em COCOON '98: Proceedings of the 4th Annual International
  Conference on Computing and Combinatorics}, pages 85--94, London, UK, 1998.
  Springer-Verlag.

\bibitem{pallo}
Jean Pallo.
\newblock An efficient upper bound of the rotation distance of binary trees.
\newblock {\em Information Processing Letters}, 73(3-4):87--92, 2000.

\bibitem{rogers}
R.~Rogers.
\newblock On finding shortest paths in the rotation graph of binary trees.
\newblock In {\em Proceedings of the Southeastern International Conference on
  Combinatorics, Graph Theory, and Computing}, volume 137, pages 77--95, 1999.

\bibitem{stt}
Daniel~D. Sleator, Robert~E. Tarjan, and William~P. Thurston.
\newblock Rotation distance, triangulations, and hyperbolic geometry.
\newblock {\em Journal of the American Mathematical Society}, 1(3):647--681,
  1988.

\end{thebibliography}

 \end{document}